\documentclass[12pt]{article} 
\title{ Rotations and $e,$ $\nu$ Propagators, Part I}  
\author{{\it Richard Shurtleff~}\thanks{affiliation and mailing 
address: Department of Applied Mathematics and Sciences, 
Wentworth Institute of Technology, 550 Huntington Avenue, 
Boston, MA, USA, ZIP 02115, telephone number: (617) 989-4338, fax 
number: (617) 989-4591 , e-mail address: shurtleffr@wit.edu}} 
\begin{document} 
          
\maketitle               
			\begin{abstract}  

Rotation symmetry is less constraining than space-time symmetry. The free electron propagator is a projection operator that we show can be constructed from rotation symmetric projection operators. Rotation-based identifications of time, space, energy, momentum, polarization matrices, and the positron hypothesis are determined by the constraints that turn rotation symmetric projection operators into the electron propagator. 

 
\end{abstract}

\pagebreak

\section{Introduction} \label{intro} 

	A free spin 1/2 particle, such as an electron, moves through space-time carrying with it an internal spin. A spin of 1/2 is only one of many choices allowed by space-time symmetry. The theorem that only integer or half integer spins are possible is well known \cite{wigner} and its agreement with observation is widely accepted.

	But isn't the logic just backwards? The rotation group of the internal spin space mimics a subgroup of the Lorentz group of space-time symmetries. One knows that going from a subgroup to the full group naturally imposes constraints in addition to the requirements of the subgroup. A moment's reflection confirms this: consider the left-right symmetric objects that can be drawn on a sheet of paper and objects with six-sided symmetry. The left-right symmetric objects can have more variety. (Also see Problem 1.) Hence it is the space-time group, the Lorentz group, that imposes more rules on its elements than the internal spin space's rotation group. Shouldn't it be space-time symmetry that is obtained by constraining spin?

	In this paper we start with the more general, the more primitive group - the rotation group of a Euclidean space. For completeness, elementary results are collected in Sec.~\ref{2d} from the well known theory of the $2 \times 2$ matrix representation of rotations in Euclidean space of three or four dimensions. Two 2-vectors make `pairs' whose properties are the focus of the work. Orthogonality and the normalization of a selected basis of pairs appears in Sec.~\ref{pairs}. In Sec.~\ref{proj} the projection operators are obtained for the basis pairs. The momentum (as-yet-undefined functions) and space occur as parameters of a delta function, which is the part of the projection operator that selects the continuous parameters of a specific basis.

	The projection operators have rotational symmetry. To achieve spacetime symmetry, we add two projection matrices and define space, time, energy, and momentum appropriately. An important step in this process, which appears in Sec.~\ref{space-time}, is making the space-time 4-vector in the delta function phase be the same as the space-time 4-vector in the two operator summed matrix. The common 4-vector is the energy-momentum with the phase and matrix typically denoted $p_{\mu} x^{\mu}$ and  $p_{\mu} \gamma^{\mu},$ respectively. Comparison with well-known formulas shows that the final two summed projection operators are just the electron and positron propagators in the absence of interactions.

	The essential constraints applied to the rotation based quantities are (i) the specific choice of basis, (ii) the definitions of energy and momentum, and (iii) the choice of projection operators to combine. The calculation follows, with some modification, a classic derivation of propagators by Feynman \cite{feynman1}. Furthermore the hypothesis that positive energy states propagate forward in time and negative energy states propagate backwards in time is explained as irreversible rotation in the Euclidean space.

	In Part II we apply the same procedure to a different basis and derive neutrino propagators. In Part III we consider arbitrary bases and show that the  $uv$ basis here, (\ref{uvpairs}), and the basis considered in Part II are the only bases that allow space time symmetric projection matrices (propagators). Also in Part III is a discussion of the 2-dimensional representations of the rotation group in a Euclidean space of four dimensions and the relationship between planes of rotation and wave functions. The two bases in Parts I and II are special because of a special property of four dimensional Euclidean space.


\section{Rotation Matrix and 2-Vectors} \label{2d} 

	In this section well known results are recalled for the 2-dimensional representations of the rotation group, which are also known as spinor representations. We consider rotations in a Euclidean space of three or four dimensions. In Part III it is shown that the process and the results imply that four dimensional Euclidean space is more suitable than three dimensional space.  

	Consider the $2 \times 2$ matrix $R$ with complex-valued elements 
\begin{equation} \label{R}
R = e^{i n^{k} \sigma^{k} \theta/2} = \sigma^{4} \cos (\theta/2) + i n^{k} \sigma^{k} \sin (\theta/2),
\end{equation}
where the three $n^{k}$ form a real 3-vector with unit magnitude, summation is assumed over $k \in$ $\{1,2,3 \},$ $\theta$ is real, and the sigma matrices are
\begin{equation} \label{sigma}
\sigma^{1} = \pmatrix{0 && 1 \cr 1 && 0} \hspace{0.3in} \sigma^{2} = \pmatrix{ 0 && -i \cr i && 0} \hspace{0.3in} \sigma^{3} = \pmatrix{ 1 && 0 \cr 0 && -1} \hspace{0.3in} \sigma^{4} = \pmatrix{ 1 && 0 \cr 0 && 1}.
\end{equation} 
The set of all matrices like $R$ forms a group under matrix multiplication. For small $\theta$ the matrix group is isomorphic to the rotation group in three dimensions with $R$ representing the rotation about the origin through an angle $\theta$ that preserves the axis in the direction of the unit 3-vector $n^{k}$ in a 3-dimensional Euclidean space, $E_{3{\mathrm{d}}}$. In a 4-dimensional Euclidean space the same $R$ can represent a rotation in another plane that intersects the first only at the origin.  

	The $2 \times 2$ matrix $R$ acts on ordered sets of two complex numbers, `2-vectors'.  One basis for 2-vectors consists of the two unit eigenvectors of $R$
\begin{equation} \label{u+u-}
u^{+} = e^{ i \alpha} \pmatrix{ \cos{(\rho/2)} \exp{(-i\phi/2)} \cr \sin{(\rho/2)} \exp{(+i\phi/2)}} \hspace{1cm} u^{-} = e^{ i \beta} \pmatrix{ - \sin{(\rho/2)} \exp{(-i\phi/2)} \cr \cos{(\rho/2)} \exp{(+i\phi/2)}},
\end{equation} 
where $\rho$ and $\phi$ are the polar and azimuthal angles for the unit 3-vector for the rotation plane
\begin{equation} \label{nk}
n^{k} = ( \sin \rho \cos \phi, \sin \rho \sin \phi, \cos \rho) 
\end{equation} 
and $\alpha$ and $\beta$ are arbitrary phases. By (\ref{R}), (\ref{sigma}), (\ref{u+u-}), and (\ref{nk}), we have
\begin{equation} \label{Ru+u-}
Ru^{+} = e^{+ i \theta /2}u^{+} \hspace{1cm} Ru^{-} = e^{- i \theta /2}u^{-}.
\end{equation} 
Call $u^{+}$ `spin up' and $u^{-}$ `spin down.' By (\ref{sigma}), (\ref{u+u-}), and (\ref{nk}), we get
\begin{equation} \label{nSIGMAu}
n^{k} \sigma^{k} u^{+} = + u^{+} \hspace{1cm} n^{k} \sigma^{k} u^{-} = -u^{-} \hspace{1cm} u^{\pm \, \dagger} \sigma^{k} u^{\pm} = \pm n^{k} \hspace{1cm} u^{\pm \, \dagger} \sigma^{4} u^{\pm} = 1.
\end{equation}
By (\ref{u+u-}) $u^{+}$ and $u^{-}$ are orthonormal, 
\begin{equation} \label{ortho,u+u-}
u^{+ \, \dagger} u^{+} = 1 \hspace{1.0cm}  u^{- \, \dagger} u^{-} = 1 \hspace{1.0cm} u^{+ \, \dagger} u^{-} = 0 \hspace{1.0cm} u^{- \, \dagger} u^{+} = 0.\end{equation}
Hence any 2-vector $u$ can be written as $u$ = $a u^{+} + b u^{-}$ where $a$ = $u^{+ \, \dagger} u$ and $b$ = $u^{- \, \dagger} u.$ The following matrix combinations prove useful,
\begin{equation} \label{u+u+dagger}
 u^{+} u^{+ \, \dagger} + u^{-}u^{- \, \dagger} = \sigma^{4} \hspace{1.0cm}   u^{+} u^{+ \, \dagger} - u^{-}u^{- \, \dagger} = n^{k} \sigma^{k} .
\end{equation}

\section{Pairs of 2-vectors} \label{pairs} 

	In this section, we seek to use the properties of pairs of 2-vectors to obtain energy, momentum and polarization matrices. The immediate justification for using pairs of 2-vectors is simply that four component wave functions are standard and they work. The properties of four dimensional Euclidean space are seen to justify such wave functions in Part III. 

	Consider the set of all ordered pairs of 2-vectors, hereafter simply called `pairs.' Thus each pair has four complex components. One basis is
 \begin{equation}	\label{uvpairs}
 u_{+}^{+} = \pmatrix{ e^{w/2} u^{+} \cr e^{-w/2} u^{+} } \hspace{1in} 
   u_{-}^{-} = \pmatrix{ e^{- w/2} u^{-} \cr e^{w/2} u^{-} },
\end{equation} 
$$
v_{+}^{+} = \pmatrix{ e^{-w/2} u^{+} \cr - e^{w/2} u^{+} } \hspace{1in} 
   v_{-}^{-} = \pmatrix{ - e^{ w/2} u^{-} \cr e^{-w/2} u^{-} }.
$$
This basis is special because in each pair the upper and lower 2-vectors are eigenvectors of $R$ with the same eigenvalue.  

	By (\ref{ortho,u+u-}) and (\ref{uvpairs}), the basis pairs are normalized to $2 \cosh (w)$ and are mutually orthogonal,
\begin{equation} \label{ortho,uvpairs}
 i^{\dagger} i = 2 \cosh w \hspace{1cm} i^{\dagger} j = 0, \hspace{1cm} i,j \in \{u_{+}^{+}, u_{-}^{-}, v_{+}^{+}, v_{-}^{-} \},
\end{equation}
where there is no sum over $i$ in the left expression and $i$ is not the same as $j$ in the middle expression. 

	Let $R_{+}^{+} u_{+}^{+}$ indicate applying the $2 \times 2$ matrix $R$ to the upper and lower 2-vectors of $u_{+}^{+}.$ By (\ref{Ru+u-}), we get a rotated basis with a common phase factor,
\begin{equation}	\label{Ruvpairs}
R_{+}^{+} u_{+}^{+} =  e^{+i\theta/2} u_{+}^{+} \hspace{0.5cm} R_{-}^{-} u_{-}^{-} =  e^{+i\theta/2} u_{-}^{-} \hspace{0.5cm} R_{+}^{+} v_{+}^{+} =  e^{+i\theta/2} v_{+}^{+} \hspace{0.5cm} R_{-}^{-} v_{-}^{-} =  e^{+i\theta/2} v_{-}^{-},
\end{equation}
where $R_{-}^{-}$ means applying $R^{-1}$ to both the upper 2-vector and the lower 2-vector. The rotated basis (\ref{Ruvpairs}) satisfies the orthogonality and normalization conditions (\ref{ortho,uvpairs}).

	There are other bases. Given $u_{+}^{+}$ and requiring an orthogonal basis, we must have $v_{+}^{+}.$ But $u_{-}^{-}$ could have the same ratio $e^{w}$ of upper to lower 2-vector as does $u^{+}_{+};$ just change $w \rightarrow$ $-w$ in $u_{-}^{-}$ and $v_{-}^{-}.$ The basis in (\ref{uvpairs}) is almost uniquely determined by requiring orthogonality and requiring that the basis gives the matrix expressions which display space-time symmetry. (See Part III.)

\section{Projection Operators for Pairs of 2-vectors} \label{proj} 

	A projection operator replicates selected parts of a quantity and destroys the rest. Given a linear combination of the rotated basis pairs, one can project out the coefficient of any one of the four basis pairs. The projection operator must select the ratio parameter $w$, the rotation plane $n^{k}$, the eigenvector $u^{+}$ or $u^{-}$, and the rotation angle $\theta.$  

	{\textit{Select the $u_{+}^{+}$ term from a sum of basis pairs}}. An arbitrary 4-component object $\psi,$ can be written as a linear combination of the basis pairs (\ref{uvpairs}),
\begin{equation} \label{psi,uv}
 \psi \equiv a  u_{+}^{+} + b u_{-}^{-} + c  v_{+}^{+} + d v_{-}^{-},
\end{equation}
where $a$ = ${u_{+}^{+}}^{\dagger} \psi/(2 \cosh{w}) .$ To make the transition to space-time symmetry below it is convenient to include a factor of $ \gamma^{4}.$ See (\ref{gA1}) for the definition of $\gamma^{4}.$ We have the projection matrix  
\begin{equation} \label{K,u++}
K \gamma^{4} \equiv \frac{1}{2 \cosh{w}} u_{+}^{+} u_{+}^{+ \: \dagger}  \hspace{1cm} K \gamma^{4} \psi = a  u_{+}^{+}
\end{equation}
The projection matrices for the other base pairs $u_{-}^{-},$ $v_{+}^{+},$ and $v_{-}^{-}$ are similar.
 
	{\textit{Select the ratio parameter $w$ and axis $n^{k}$}}. By (\ref{u+u-}), (\ref{nk}), and (\ref{uvpairs}), the basis pairs are functions of the ratio parameter $w$ and the unit 3-vector $n^{k}.$ Since ${n^{k}}^{2}$ = 1, in the set of four variables $w$ and $n^{k}$ only three are independent and we can use a three dimensional delta function to select specific values of these variables. However, it is not conventional to write the delta functions of $w$ and $n^{k},$ instead the delta functions are written in terms of the `momentum' $p^{j},$ which are three functions of $w$ and $n^{k},$ $p^{j}(w,n^{k}).$ This means that $w $ and $n^{k}$ are determined by $p^{k},$ i.e. $w(p^{j})$ and $n^{k}(p^{j}),$ and we can use delta functions for the three momentum components, $p^{j}-{p^{j}}^{\prime},$ instead of delta functions for $w - w^{\prime}$ and $n^{k} - {n^{k}}^{\prime}.$ The momentum function $p^{j}$ remains undetermined for now until (\ref{pk}) and (\ref{pk2}) in Sec.~\ref{space-time} where the requirements of space-time symmetry are considered.

	Any function of $p^{j},$ such as the basis pair $u_{+}^{+}$ = $u_{+}^{+}(p^{j}),$ can be made the amplitude of a plane wave,
\begin{equation}	\label{psi,f(p)}
\psi_{p}(1) \equiv e^{ip^{k} x_{1}^{k}} u_{+}^{+}(p^{j}).
\end{equation}
Call the phase $p^{k} x^{k}$ the `plane wave phase.' By integrating with the three dimensional delta function, we get
\begin{equation}	\label{Kpsi,f(p)}
  \int d^{3} x_{1} K_{p^{\prime}}(2,1) \psi_{p}(1) =  e^{ip^{k} x_{2}^{k}} u_{+}^{+}(p^{j}) = \psi_{p}(2),
\end{equation}
where 
\begin{equation}	\label{K,f(p)}
K_{p^{\prime}}(2,1) \equiv \int \frac{d^3 p^{ \prime} }{(2 \pi)^3} e^{i p^{k \, \prime} x_{2}^{k}} e^{- i p^{k \, \prime} x_{1}^{k}}.
\end{equation} 
By (\ref{Kpsi,f(p)}) the delta function has no effect on the plane wave except to reparameratize the phase, i.e. $x_{1}^{k} \rightarrow$ $x_{2}^{k}.$

	{\textit{Rotation angle}} $\theta$. Rotation of $u_{+}^{+}$ or $v_{+}^{+}$ by $R$ through an arbitrary angle $\theta$ produces a phase factor $\psi_{\theta}$ = $e^{i \theta/2}$ by (\ref{Ruvpairs}). The arbitrary phase can be replicated by multiplication with another phase factor $K_{\theta}(2,1),$ 
\begin{equation}  \label{psi,theta}
\psi_{\theta}(1) \equiv e^{i\theta_{1}/2} \hspace{2cm}
K_{\theta}(2,1) \equiv e^{i\theta_{2}/2 } e^{-i\theta_{1}^{\prime}/2}
\end{equation}
\begin{equation}  \label{Kpsi,theta}
K_{\theta}(2,1) \psi_{\theta}(1) = (e^{i\theta_{2}/2 } e^{-i\theta_{1}^{\prime}/2}) e^{i\theta_{1}/2}
=  e^{i(\theta_{2} - \theta_{1}^{\prime} + \theta_{1})} = e^{i\theta_{2}/2} = \psi_{\theta}(2).
\end{equation} 
There is no loss of generality in choosing $\theta_{1}$ = $\theta_{1}^{\prime}$ because $\theta_{2}$ can have any real value and therefore represent any angle in the new function $\psi_{\theta}(2)$ = $e^{i \theta_{2} / 2}$, just as $\theta_{1}$ can be any angle in the original $\psi_{\theta}(1).$ 

	The plane wave phase $ p^{k} x^{k}$ differs from the eigenvalue phase $\theta/2.$ Let $\Delta$ be the difference so that 
\begin{equation} \label{theta,Delta,px}
\frac{\theta}{2} = -\Delta + p^{k} x^{k}.
\end{equation}
Thus we can write $\psi_{\theta}(1)$ and $K_{\theta}$ as 
\begin{equation}	\label{psi,Delta,p}
\psi_{\theta}(1) = e^{-i \Delta_{1}} e^{ i p^{k} x_{1}^{k}} \hspace{2cm} 
K_{\theta}(2,1) = e^{-i \Delta_{2}^{\prime}} e^{ i p^{k \, \prime} x_{2}^{k}} e^{i \Delta_{1}^{\prime}} e^{  -i p^{k \, \prime} x_{1}^{k}} .
\end{equation}
Since the rotation angle $\theta$ can have any real value, by (\ref{theta,Delta,px}) the difference $\Delta$ can have any real value at each point $x^{k}.$

	{\textit{Projection Operators}}. Now we project the positive or negative ratio, spin up or spin down portion out of an arbitrary 4-component wave function $\psi$ by combining the various  $4 \times 4$ projection matrices, delta functions, and phase adjustments.  

	Expand the wave function $\psi(1)$ over the basis pairs (\ref{Ruvpairs}). 
\begin{equation}	\label{psi(1),theta}
\psi(1) = a R_{+}^{+} u_{+}^{+} + bR_{-}^{-} u_{-}^{-} + cR_{+}^{+} v_{+}^{+} + dR_{-}^{-} v_{-}^{-} =
\end{equation}  
$$
 =  e^{i \theta_{1} / 2}(a  u_{+}^{+}  + b  u_{-}^{-} + c  v_{+}^{+}  + d  v_{-}^{-}) = e^{i \theta_{1} / 2} \psi_{0}
$$
where $a$ = ${(R_{+}^{+} u_{+}^{+})}^{\dagger} \psi / (2 \cosh{w})$ and $\psi_{0}$ is $\psi(1)$ for $\theta_{1}$ = 0. The $R_{+}^{+} u_{+}^{+}$ projection operator combines the $K$s from (\ref{K,u++}), (\ref{psi,theta}), and (\ref{psi,Delta,p}). We have
\begin{equation}	\label{K(2,1)}
  K(2,1,R_{+}^{+} u_{+}^{+}) \gamma^{4} \equiv \int \frac{d^3 p^{\prime} }{(2 \pi)^3} \frac{1}{2 {\cosh{(w)}}^{\prime}}e^{i \theta_{2}^{\prime}}  u_{+}^{+ \, \prime} {u_{+}^{+ \, \prime }}^{\dagger} e^{ -i \theta_{1}^{\prime}/2} ,
\end{equation} 
where $\theta_{i}^{\prime}$ = $- \Delta_{i}^{\prime} + p^{k \, \prime} x_{i}^{k}.$  By (\ref{K,u++}), (\ref{Kpsi,f(p)}), and (\ref{Kpsi,theta}), we get 
\begin{equation}	\label{K(2,1)psi}
 \int d^{3} x_{1} K(2,1,R_{+}^{+} u_{+}^{+}) \gamma^{4} \psi(1) = a e^{i \theta_{2} / 2} u_{+}^{+} = aR_{+}^{+} u_{+}^{+} = \psi(2)_{b=c=d=0},
\end{equation} 
where $ \theta_{2}/2 $ = $- \Delta_{2}^{\prime} + \Delta_{1}^{\prime} - \Delta_{1} + p^{k } x_{2}^{k} $ = $- \Delta_{2} + p^{k } x_{2}^{k},$ i.e. we make $\Delta_{2}$ obey $\Delta_{2} - \Delta_{1}$ = $\Delta_{2}^{\prime} - \Delta_{1}^{\prime}.$  Equation (\ref{K(2,1)psi}) illustrates a projection operator at work: $K(2,1,R_{+}^{+} u_{+}^{+}) \gamma^{4}$ projects out the $R_{+}^{+} u_{+}^{+}$ portion of $\psi(1).$ The projection operators for the other basis pairs, $K(2,1,R_{-}^{-}u_{-}^{-})\gamma^{4},$ $K(2,1,R_{+}^{+}v_{+}^{+})\gamma^{4},$ $K(2,1,R_{-}^{-}v_{-}^{-})\gamma^{4}$ are given by the expression (\ref{K(2,1)}) with the basis pair $R_{+}^{+} u_{+}^{+}$ replaced by $R_{-}^{-}u_{-}^{-},$ $R_{+}^{+}v_{+}^{+},$ and $R_{-}^{-}v_{-}^{-},$ respectively.

	The projection operator $K(2,1,R_{+}^{+} u_{+}^{+})$ in (\ref{K(2,1)}) is invariant under rotations when both $p^{k}$ and $x^{k}$ transform as vectors leaving $p^{k} x^{k}$ and $\theta$ invariant. To show that $u_{+}^{+ \, \prime} {u_{+}^{+ \, \prime }}^{\dagger}$ is a rotation invariant, see (\ref{u++u++g}) below. Making spacetime invariant projection operators using the four rotation invariant projection operators, like $K(2,1,R_{+}^{+} u_{+}^{+}),$ is a task for the next section.

\section{Space-Time Symmetry, the Electron Propagator} \label{space-time} 

	In this section two projection operators are obtained and written in a way that makes their invariance under space-time transformations evident. For phases, we show how the quantity $\Delta$ in (\ref{theta,Delta,px}) becomes $Et,$ where $E$ is the `energy' and $t$ is time. We see what it means to require that positive energy states must propagate forward in time and negative energy states must propagate backwards in time, i.e. the positron hypothesis is explained. For matrices, we find that we can combine the two projection operators for the positive ratio pairs $u_{+}^{+}$ and $u_{-}^{-},$ thereby making a positive energy matrix. Then the two negative ratio operators combine to make a negative energy matrix.

	{\textit{Time}}. Time can be introduced by applying an integral expression \cite{feynman2} for the phase factor $e^{-i\Delta}$ ,
\begin{equation} \label{expiDelta}	
e^{-i \Delta} = \frac{i}{ \pi} \int_{-\infty}^{\infty} da \frac{e^{-ia\Delta}}{a^2-1+i\epsilon}
\end{equation}
where the value of $\Delta$ is positive and $\epsilon$ is small and positive so that the pole at $a$ = 1 is included when the integral is evaluated by contour integration.  For brevity, we do not consider $\Delta <$ 0 except to note that, for $\Delta_{2}^{\prime} < 0$, change the overall sign and change the sign of $\epsilon$ in (\ref{expiDelta}).

	By (\ref{theta,Delta,px}), (\ref{K(2,1)}), and (\ref{K(2,1)psi}), we have
$$ \psi(2)_{b=c=d=0} = \int d^3x_{1}K(2,1,R_{+}^{+}u_{+}^{+}) \gamma^{4} \psi(1)$$ $$ =  \int d^3x_{1}\int \frac{d^3 p^{\prime} }{(2 \pi)^3} \frac{1}{2 {\cosh{(w)}}^{\prime}}e^{i \theta_{2}^{\prime}/2}  u_{+}^{+ \, \prime} {u_{+}^{+ \, \prime }}^{\dagger} e^{ -i \theta_{1}^{\prime}/2}e^{i \theta_{1} / 2} \gamma^{4} \gamma^{4} \psi_{0}
$$
\begin{equation}	\label{K++K--psi}
= \int d^3 x_{1}\lbrack \int \frac{d^3 p^{\prime} }{(2 \pi)^3} \frac{1}{2 {\cosh{w}}^{\prime}}e^{i \theta_{2}^{\prime}/2}  u_{+}^{+ \, \prime} {u_{+}^{+ \, \prime }}^{\dagger} \gamma^{4}\rbrack e^{ i (\Delta_{1}^{\prime}- \Delta_{1})} e^{i (p^{k \; \prime} - p^{k})x_{1}^{k}}  \gamma^{4} \psi_{0}.
\end{equation}
Let $I$ be the quantity in (\ref{K++K--psi}) in brackets. By (\ref{theta,Delta,px}) and (\ref{expiDelta}) and for $\Delta_{2}^{\prime} >$ 0, we get
$$ I = \int \frac{d^3 p^{\prime} }{(2 \pi)^3} \frac{1}{2 {\cosh{w}}^{\prime}} e^{-i\Delta_{2}^{\prime}} e^{ip^{k \; \prime}x_{2}^{k}} u_{+}^{+ \, \prime} {u_{+}^{+ \, \prime }}^{\dagger} \gamma^{4} 
$$
$$= \int \frac{d^3 p^{\prime} }{(2 \pi)^3} \frac{1}{2 {\cosh{w}}^{\prime}}  ( \frac{i}{ \pi}  \int_{-\infty}^{\infty} da \frac{e^{-ia\Delta_{2}^{\prime}}}{a^2-1+i\epsilon}) e^{ip^{k \; \prime}x_{2}^{k}} u_{+}^{+ \, \prime} {u_{+}^{+ \, \prime }}^{\dagger} \gamma^{4} 
$$
\begin{equation} \label{QED}
=  i  \int \frac{d^4 p^{\prime} }{(2 \pi)^4} e^{-i(p^{4 \; \prime}t_{2} - p^{k \; \prime}x_{2}^{k})}  \frac{ (\pm m) u_{+}^{+ \, \prime} {u_{+}^{+ \, \prime }}^{\dagger} \gamma^{4}}{{p^{4 \; \prime}}^2- m^{2} \cosh^{2}{w}^{\prime} + i\epsilon} 
\end{equation}
where we introduce a fourth component of momentum and a fourth component of displacement, i.e. energy and time,
\begin{equation}	\label{p4t}
{p^{4}}^{\prime} \equiv \pm m {\cosh{(w)}}^{\prime} a   \hspace{1cm}  a \Delta_{2}^{\prime} = {p^{4}}^{\prime} t_{2} , 
\end{equation}
where $m$ is a positive constant tentatively called the `bare mass.' 

	{\textit{Surface Integral}}. Next, in (\ref{K++K--psi}) identify the integral over the 3-space $x_{1}^{k}$ with a surface integral in space-time,
\begin{equation}	\label{K++K--psi1}
\psi(2)_{b=c=d=0} = \int_{S} d^4 x_{1}I e^{ i (\Delta_{1}^{\prime}- \Delta_{1})} e^{i (p^{k \; \prime} - p^{k})x_{1}^{k}} N_{\mu} \gamma^{\mu} \psi_{0}
\end{equation}
where $x^{4}$ = $t,$ $\mu \in$ $\{1,2,3,4\},$ $N^{\mu}$ = $\{N^{k},N^{4}\}$ is the unit normal to the three dimensional surface of integration $S$ in four dimensional space-time, and the space-time summation convention is used, $N_{\mu} \gamma^{\mu}$ = $N^{4} \gamma^{4} - N^{k} \gamma^{k}.$ In (\ref{K(2,1)psi}) and (\ref{K++K--psi}), the integration is over the three dimensional surface $x_{1}^{k}$ in the special space-time reference frame with the normal in the time direction,  $N^{\mu}$ = $\{0,0,0,1\}.$ 

	{\textit{Space-time Transformation}}. The linear transformations of $p^{\mu}$ and $x^{\mu}$ that preserve the sum $p_{\mu} x^{\mu}$ = ${p^{4}}t - {p^{k}}x^{k}$ are the space-time transformations, i.e. Lorentz transformations. Only homogeneous transformations are considered here. We denote the coefficients of one such transformation by $\Lambda_{\nu}^{\mu},$
\begin{equation}	\label{lambda}
 P^{\, \mu} = \Lambda_{\nu}^{\mu} p^{\nu} \hspace{1cm} P_{\mu}P^{\mu} = {P^{\,4}}^{2} - {P^{\, k}}^{2}  = {p^{4}}^{2} - {p^{k}}^{2} .
 \end{equation}
Thus the phases, and hence also the rotation angle $\theta,$ in (\ref{QED}) are space-time invariants because they contain scalar products such as ${p_{\mu}}^{\prime}{x^{\mu}}^{\prime}.$ 

	{\textit{Matrices}}. The projection operator $K(2,1,R_{+}^{+} u_{+}^{+}) \gamma^{4}$ rewritten with (\ref{K++K--psi}) and (\ref{QED}) still fails to have space-time symmetry because of the quantity $m u_{+}^{+ \, \prime} {u_{+}^{+ \, \prime }}^{\dagger} \gamma^{4}$ in (\ref{QED}). We can express the matrix $m u_{+}^{+ \, \prime} {u_{+}^{+ \, \prime }}^{\dagger} \gamma^{4}$ as a sum of sixteen linearly independent $4 \times 4$ matrices such as the set of gamma matrices in Appendix A. By (\ref{u+u+dagger}), (\ref{uvpairs}), and (\ref{M1}), we get 
\begin{equation}	\label{u++u++g}
m u_{+}^{+ } {u_{+}^{+  }}^{\dagger} \gamma^{4} = \frac{1}{2}[m \cdot 1 - m \sinh{(w)} n^{i} \gamma^{i} + m\cosh{(w)} \gamma^{4}] + \hspace{4cm}
\end{equation}
$$+\frac{i}{2}[ m\cosh{(w)} n^{j} \gamma^{j} \gamma^{5} - m \sinh{(w)}\gamma^{4} \gamma^{5} + m n^{k} \gamma^{5} \gamma^{4} \gamma^{k}],$$
where we drop the primes. 

	Let $n^{k}$ and $\gamma^{k}$ transform as 3-vectors under rotations. One can show that $\gamma^{5}$, (\ref{gA1}), is invariant under rotations. Thus the matrix (\ref{u++u++g}) is a rotation invariant. While the matrix (\ref{u++u++g}) is a rotation invariant, it is not a space-time invariant; see Problem 3.

	The expansion of $ m {u_{-}^{-}} {u_{-}^{- }}^{\dagger} \gamma^{4}$ over the same set of gammas gives
\begin{equation}	\label{u--u--g}
m u_{-}^{- } {u_{-}^{-  }}^{\dagger} \gamma^{4} = \frac{1}{2}[m \cdot 1 - m \sinh{(w)} n^{i} \gamma^{i} + m\cosh{(w)} \gamma^{4}] + \hspace{4cm}
\end{equation}
$$-\frac{i}{2}[  m\cosh{(w)} n^{j} \gamma^{j} \gamma^{5} - m \sinh{(w)}\gamma^{4} \gamma^{5} + m n^{k} \gamma^{5} \gamma^{4} \gamma^{k}].$$
When we add the two expressions, we get
\begin{equation}	\label{Epmatrix}
m ( u_{+}^{+ } {u_{+}^{+  }}^{\dagger} + u_{-}^{- } {u_{-}^{- }}^{\dagger} ) \gamma^{4} = m \cosh{(w)} \gamma^{4} - m \sinh{(w)} n^{k} \gamma^{k} + m \cdot 1  .
\end{equation}
Let $\gamma^{k}$ and $\gamma^{4}$ form a 4-vector, i.e. transform as in (\ref{lambda}). See \cite{feynman3} for why this can be allowed. Let $m$ and the unit matrix $1$ be scalars under space-time transformations. If $m \sinh{(w)} n^{k}$ and $m \cosh{w}$ also transform as the components of a 4-vector, then the expression (\ref{Epmatrix}) is a space-time invariant. 

	{\textit{Energy-momentum}}. We now make the choice of momentum functions and, at the same time, we choose the energy function. Note that both $\{m \sinh{(w)} n^{k}, m \cosh{w} \}$ and $\{p^{k}, p^{4}\}$ = $\{p^{k}, \pm m \cosh{(w)} a\}$ are 4-vectors. We define the momentum functions
\begin{equation} \label{pk}
p^{k} \equiv m \sinh{(w)} n^{k} \hspace{1cm} p^{4} = +m \cosh{(w)} a.
\end{equation}
Then the pole in (\ref{expiDelta}) occurs when $p^{4}$ has the value $E,$
\begin{equation}	\label{E}
E \equiv m \cosh{w} \hspace{1cm} \Delta = Et.
\end{equation} 
The energy-momentum components connect the invariant phase $\theta/2$ = $Et$ $-p^{k} x^{k}$ and the matrix (\ref{Epmatrix}), which we can now rewrite as
\begin{equation}	\label{Epmatrix1}
m ( u_{+}^{+ } {u_{+}^{+  }}^{\dagger} + u_{-}^{- } {u_{-}^{- }}^{\dagger} ) \gamma^{4} = E \gamma^{4} - p^{k} \gamma^{k} + m \cdot 1  .
\end{equation}

	The phases $\Delta_{1}^{\prime}- \Delta_{1}$ and $(p^{k \; \prime} - p^{k})x_{1}^{k}$ in (\ref{K++K--psi}) can be treated similarly. Since the propagated wave function $\psi(2)$ is the initial wave function for the next propagator, the phase for $\psi(1)$ should have the same form as $\psi(2).$ It follows that, as in (\ref{p4t}), we can write $\Delta_{1}^{\prime}$ = $E^{\prime} t_{1}$ and $\Delta_{1}$ = $E t_{1}$ so that the integral over $t_{1}$ = $x_{1}^{4}$ makes a delta function $\delta(E^{\prime}- E).$ In the special reference frame of Sec.~\ref{proj}, the basis pairs are functions of $p^{k},$ i.e. $w$ and $n^{k},$ so the delta function for time does not constrain the basis pairs. But in a more general frame the four dimensional delta function is needed to select the basis pairs corresponding to the $w$ and $n^{k}$ of the special frame.

	$E \geq m$ {\textit{Propagator}}. The space-time symmetry of the sum of the two positive ratio operators, $K(2,1,R_{+}^{+} u_{+}^{+}) \gamma^{4}$ and $K(2,1,R_{-}^{-} u_{-}^{-}) \gamma^{4},$ is evident when the above results are combined. By (\ref{K++K--psi}), (\ref{QED}), (\ref{K++K--psi1}), (\ref{pk}), (\ref{E}), and (\ref{Epmatrix1}), we get
\begin{equation}	\label{QED2}
\psi(2)_{c=d=0} = \int d^3x_{1}(K(2,1,R_{+}^{+}u_{+}^{+}) + K(2,1,R_{-}^{-}u_{-}^{-})) \gamma^{4} \psi(1) =
\end{equation}
$$ = i\int \frac{d^4 x_{1} d^4 p^{\prime} }{(2 \pi)^4} e^{-ip_{\mu}^{\prime}x_{2}^{\mu}}  \frac{ p_{\nu}^{\prime} \gamma^{\nu} + m \cdot 1}{p_{\eta}^{\prime}p^{\eta \, \prime}  - m^2 + i\epsilon} e^{ i p_{\tau}^{\prime} x_{1}^{\tau}} N_{\sigma} \gamma^{\sigma} \psi(1),$$
where we used another integral formula,
\begin{equation} \label{expiDelta2}	
e^{-i \Delta} =  \frac{i}{ \pi} \int_{-\infty}^{\infty} da \frac{ae^{-ia\Delta}}{a^2-1+i\epsilon},
\end{equation}
 to replace $E \rightarrow$ $Ea$ = $p^{4}$ in the matrix $(E \gamma^{4} - p^{k} \gamma^{k} + m \cdot 1).$ 

	Spacetime symmetry is evident in (\ref{QED2}) because scalar products like $p_{\mu}p^{\mu }$ are invariant under space-time transformations and the integrals are over four dimensions.

	{\textit{Experimental confirmation}}. The fraction in (\ref{QED2}) is sometimes written $(p_{\mu}^{\prime} \gamma^{\mu} - m)^{-1}$ because
\begin{equation}	\label{m-1}
(p_{\mu}^{\prime} \gamma^{\mu} - m)(p_{\nu}^{\prime} \gamma^{\nu} + m) = (p_{\sigma}^{\prime} {p^{\sigma}}^{\prime}- m^2) .
\end{equation}
The projection operator (\ref{QED2}) can be shown to be the Green's function for the inverted matrix operator $(p_{\mu}^{\prime} \gamma^{\mu} - m).$ \cite{bjorken} And one can show that the Green's function is related to the Dirac equation. This justifies the use of the terms `energy' and `momentum' for $E$ and $p^{k}$ because the formulas have been used to describe electrons and other spin 1/2 particles. In particular, expression (\ref{QED2}) is equivalent to the standard electron propagator for positive energy states.

	$E \leq -m$ {\textit{Propagator}}. For the basis pairs $v_{+}^{+}$ and $v_{-}^{-}$ we can follow the steps that lead to (\ref{Epmatrix}). We have
\begin{equation}	\label{Epmatrix2}
m ( v_{+}^{+ \, \prime} {v_{+}^{+ \, \prime }}^{\dagger} + v_{-}^{- \, \prime} {v_{-}^{- \, \prime }}^{\dagger} ) \gamma^{4} = -(m \cdot 1 - m \sinh{(w)} n^{k} \gamma^{k} - m \cosh{(w)} \gamma^{4}) .
\end{equation}
Then, for $\Delta_{2}^{\prime} >$ 0, we get
$$\psi(2)_{a=b=0} \hspace{.5cm} = \int d^3x_{1}[K(2,1,R_{+}^{+}v_{+}^{+}) + K(2,1,R_{-}^{-}v_{-}^{-})] \gamma^{4} \psi(1) = $$
$$
= -\int d^3x_{1}\int \frac{d^3 p^{\prime} }{(2 \pi)^3} \frac{1}{2 m \cosh{w}^{\prime}} e^{-i\Delta_{2}^{\prime}} e^{ip^{k \; \prime}x_{2}^{k}} ( -m \cosh{(w^{\prime})} \gamma^{4} - m \sinh{(w^{\prime})} n^{k \; \prime} \gamma^{k} + m) \gamma^{4} \psi(1)
$$
\begin{equation}	\label{QED3}
= i\int \frac{d^4 x_{1} d^4 p^{\prime} }{(2 \pi)^4} e^{-ip_{\mu}^{\prime}x_{2}^{\mu}}  \frac{ p_{\nu}^{\prime} \gamma^{\nu} + m}{p_{\tau}^{\prime}p^{\tau \, \prime}  - m+i\epsilon} e^{ i p_{\eta}^{\prime} x_{1}^{\eta}} N_{\sigma} \gamma^{\sigma} \psi(1)
\end{equation}
Since the $\cosh{w^{\prime}}$ term is negative in the matrix factor $( -m \cosh{(w^{\prime})} \gamma^{4} - m \sinh{(w^{\prime})} n^{k \; \prime} \gamma^{k} + m),$ we need to choose the negative sign in (\ref{p4t}) and we get new definitions for ${p^{4}}^{\prime}$ and $t_{2}$ in (\ref{QED3}). We define new momentum and energy functions
\begin{equation} \label{pk2}
p^{k} = m \sinh{(w)} n^{k} \hspace{1cm} p^{4} = -m \cosh{(w)} a.
\end{equation}
Note that we must choose the negative sign in the $\pm m$ term in (\ref{QED}) and (\ref{p4t}). The pole in (\ref{expiDelta}) occurs when $p^{4}$ has the value $E,$
\begin{equation}	\label{E2}
E \equiv -m \cosh{w} \hspace{1cm} \Delta = Et.
\end{equation} 

	{\textit{Positron Hypothesis}}. We now consider the positron hypothesis, i.e. positive energy states propagate forward in time while negative energy states propagate backwards in time. By (\ref{theta,Delta,px}), (\ref{pk}), and (\ref{E}), we get an expression for time $t$ for positive energy states 
\begin{equation}	\label{poshyp}
t = \frac{1}{\cosh{(w)}} (\sinh{(w)}n^{k}x^{k} - \frac{\theta}{2m}) \hspace{1cm} (E \geq m > 0).
\end{equation}
And, by (\ref{theta,Delta,px}), (\ref{pk2}), and (\ref{E2}), we get an expression for time $t$ when the energy is negative, 
\begin{equation}	\label{neghyp}
t = \frac{1}{\cosh{(w)}} (\frac{\theta}{2m} - \sinh{(w)}n^{k}x^{k} ) \hspace{1cm} (E \leq m < 0).
\end{equation}
In the special reference frame of Sec.~\ref{proj} let the time $t_{1}$ be zero on the three dimensional surface of integration $x_{1}^{k}.$ Identify $x_{2}^{k}$ with $x_{1}^{k}$ for $t_{2}$ = 0. By (\ref{poshyp}), if $\theta$ decreases then $t_{2}$ is positive everywhere, $t_{2} >$ 0 for positive energy. And, and by (\ref{neghyp}), if $\theta$ decreases then $t_{2}$ is negative everywhere, $t_{2} >$ 0 for negative energy. We have, at any point $x_{2}^{k},$
$$
 {\mathrm{positive}}  \hspace{0.25 cm} {\mathrm{energy:}} \hspace{0.3 cm} E \geq m \hspace{0.3 cm} {\mathrm{and}} \hspace{0.3 cm} \theta \hspace{0.15 cm} {\mathrm{decreasing}} \hspace{0.3cm} \Rightarrow \hspace{0.3cm}  t_{2} > 0  \hspace{0.3 cm}	{\mathrm{and}} \hspace{0.3 cm} t_{2} \hspace{0.15 cm} {\mathrm{increasing}} \hspace{1. cm}
$$
\begin{equation}	\label{t+-a}
 {\mathrm{negative}}  \hspace{0.25 cm} {\mathrm{energy:}} \hspace{0.3 cm} E \leq -m \hspace{0.3 cm} {\mathrm{and}} \hspace{0.3 cm} \theta \hspace{0.15 cm} {\mathrm{decreasing}} \hspace{0.3cm} \Rightarrow \hspace{0.3cm}  t_{2} < 0  \hspace{0.3 cm}	{\mathrm{and}} \hspace{0.3 cm} t_{2} \hspace{0.15 cm} {\mathrm{decreasing}} .
\end{equation}
Thus the positron hypothesis is equivalent to requiring that the rotation angle $\theta$ decreases with time, i.e. the rotation angle always decreases as the absolute value of the time increases. 

	{\textit{Remarks}}. One should note that the functions chosen for $p^{k}$ and $E$ in (\ref{pk}), (\ref{E}), (\ref{pk2}), and (\ref{E2}) have immediate impact on the properties of the space and time variables $x^{\mu}$ with which they appear in the delta function phase.

	Since the $uv$ basis (\ref{uvpairs}) and $\psi$ are originally written as functions of $w$ and $n^{k}$ and the delta functions are in terms of $p^{\mu},$ the two different definitions of energy (\ref{E}) and (\ref{E2}) require the $uv$ basis and $\psi$ to be two different functions of $p^{4}.$

	In this part, we have obtained the standard QED propagator for a free spin 1/2 particle moving forward in time, (\ref{QED2}), and the QED propagator for a free spin 1/2 particle moving backward in time, (\ref{QED3}). The propagators are expressed in terms of quantities defined on a rotation group in a Euclidean space. 

\appendix

\section{Sixteen Gammas}

	The set of $4 \times 4$ matrices is spanned by a basis consisting of sixteen linearly independent $4 \times 4$ matrices, which can be chosen in many ways. We choose a convenient set of matrices, called the `gammas', to express the results in the text simply. \cite{messiah}

	Start by defining five gammas in terms of the sigmas (\ref{sigma}),
\begin {equation} \label{gA1}
  \gamma^{k} = \pmatrix{ 0 && - \sigma^{k} \cr \sigma^{k} && 0} \hspace{1cm} \gamma^{4} = \pmatrix{ 0 && \sigma^{4} \cr \sigma^{4} && 0} \hspace{1cm} \gamma^{5} = \gamma^{4} \gamma^{1} \gamma^{2} \gamma^{3},
\end{equation}
where $k \in$ $\{1,2,3\}.$ Then the set of sixteen gammas is
\begin {equation} \label{gA2}
 \gamma_{A} = \{1; \hspace{0.3cm} \gamma^{k};  \hspace{0.3cm}  \gamma^{4}; \hspace{0.3cm}  \gamma^{k} \gamma^{4}; \hspace{0.3cm}  \gamma^{5} \gamma^{4} \gamma^{k}; \hspace{0.3cm} \gamma^{4} \gamma^{5} ; \hspace{0.3cm}  \gamma^{k}\gamma^{5}; \hspace{0.3cm}  \gamma^{5} \},
\end{equation}
where $1$ is the unit matrix and $A$ takes successive integer values starting with $\gamma_{0}$ = $1$ to $\gamma_{15}$ = $\gamma^{5}.$

	The square of each matrix (\ref{gA2}) is either plus or minus the unit matrix and the trace of each matrix is zero except for $\gamma_{0}.$ We have
\begin {equation} \label{gA3}
  {\gamma_{A}}^2 = \epsilon^{A} \cdot 1  \hspace{1cm} {\mathrm{trace }}(\gamma_{0}) = {\mathrm{trace }}(1) =4 \hspace{1cm} {\mathrm{trace }}(\gamma_{A}) \mid_{A \neq 0} \hspace{0.1cm}= 0 ,
\end{equation}
where $\epsilon^{A}$ = $\pm 1.$ We can use these properties to obtain the coefficients $\alpha^{A}$ in the expansion of a $4 \times 4$ matrix $M$. We get
\begin {equation} \label{M1}
  M = \alpha^{A}{\gamma^{A}}   \hspace{1cm} \alpha^{A} = \frac{\epsilon^{A}}{4} {\mathrm{trace }}(\gamma_{A}.M),
\end{equation}
where $A$ is summed in the left expression but $A$ is not summed in the expression on the right.

 \section{Problems} 

\noindent 1. Consider the set $S$ of all spheres centered on the origin of a given inertial frame $F.$ Each sphere is invariant under rotations about the origin in $F.$ What subset $s$ of $S$ contains spheres that are also spheres in the sets $S^{\prime}$ of spheres in all other space-time reference frames $F^{\prime}?$ [Hence space-time symmetry is more restrictive than rotational symmetry.]

\vspace{0.3cm}
\noindent 2. Verify (\ref{expiDelta}).

\vspace{0.3cm}
\noindent 3. (a) Derive (\ref{u++u++g}). (b) Show that $m u_{+}^{+} {u_{+}^{+  }}^{\dagger} \gamma^{4}$ is invariant under rotations and also under boosts in the direction of $p^{k},$ but not invariant under boosts perpendicular to $p^{k}.$ Assume that $\gamma^{\mu}$ transforms as a 4-vector, i.e. like $p^{\mu}$ in (\ref{lambda}).

\vspace{0.3cm}
\noindent 4. The proper time $\tau$ is the ordinary time $t$ in a space-time reference frame with $p^{k}$ = 0. How does the proper time along a path $x^{\mu}$ that is proportional to $p^{\mu}$ depend on the rotation angle $\theta?$

\vspace{0.3cm}
\noindent 5. Let $A$ = $(u_{+}^{+ } {u_{+}^{+  }}^{\dagger} + u_{-}^{- } {u_{-}^{- }}^{\dagger})/(2 \cosh{w})$ and $B$ = $(v_{+}^{+ } {v_{+}^{+  }}^{\dagger} + v_{-}^{- } {v_{-}^{- }}^{\dagger})/(2 \cosh{w}).$ (a) Show that $A$ and $B$ are idempotent and divisors of zero, i.e. $A^2$ = $A,$ $B^2$ = $B,$ and $AB$ = 0. Also show that $A + B$ = 1. (b) Then use   (\ref{pk}), (\ref{E}), and (\ref{Epmatrix1}) and (\ref{Epmatrix2}), (\ref{pk2}), and (\ref{E2}) to rewrite $A$ and $B$ in terms of $E,$ $p^{k},$ and $\gamma^{\mu}.$

\begin{tabular}{rl}
\multicolumn{2}{c}{\rule[-3mm]{0mm}{8mm}\bfseries Glossary}\\
\multicolumn{2}{c}{\rule[-3mm]{0mm}{8mm}Euclidean Space Rotations (Internal Spin Space)}\\
$R$ & $2 \times 2$ rotation matrix  \\
$\theta$ & rotation angle  \\
$n^{k}$ & unit 3-vector \\
$u^{+},$ $u^{-}$ & eigenvectors of $R$ with eigenvalues $e^{+i \theta/2},$ $e^{-i \theta/2}$ \\
$u_{+}^{+},$ $u_{-}^{-},$ $v_{+}^{+},$ $v_{-}^{-}$ & pairs of eigenvectors; a basis for the set of all pairs  \\
$e^{+w},$ $e^{-w}$ & ratio of upper to lower 2-vector in a pair \\ 
\multicolumn{2}{c}{\rule[-3mm]{0mm}{8mm}Space-time Quantities}\\
$m$ & mass is not defined in terms of rotation quantities \\
$E$ = $m \cosh{w}$ & energy  \\
$p^{k}$ = $ m \sinh{(w)} n^{k}$ & momentum \\
$x^{k}$ & space occurs as parameters in a 3d delta function \\
 & that selects $w$ and $n^{k}$ \\
$\tau$ = $\theta/(2m)$ & proper time  \\
$Et$ = $ p^{k}x^{k} - m\theta/2$ & time arises from the difference between \\
 & delta function phase and eigenphase \\
$\psi$ & wave function for the state with definite \\
	& momentum and energy, i.e. definite $w$ and $n^{k}$  \\
$K \gamma^{4},$ $K_{p^{\prime}},$ $K_{\theta},$ $K(2,1,R_{+}^{+} u_{+}^{+}) \gamma^{4}$ & parts of projection operators used \\ 
 & to select and replicate various parts of $\psi$ 
\end{tabular}


\begin{thebibliography}{9}
\bibitem{wigner} Wigner, E., Annals of Mathematics, 40, 149 (1937). See also, for example, Weinberg, S., The Quantum Theory of Fields I (Cambridge University Press, Cambridge, 1995), Sec.~2.5 and references therein.
\bibitem{feynman1} Feynman, R. P., Quantum Electrodynamics(Addison-Wesley, Redwood City, California, 1961), Fifteenth through Seventeenth Lectures.
\bibitem{pairs} Shurtleff, R., quant-ph/9903093, and references therein.
\bibitem{feynman2} Feynman, R., op cit., p. 85.
\bibitem{feynman3} For example, Feynman, R., op cit., p. 45.
\bibitem{bjorken} See, for example, Bjorken, J. D. and Drell, S. D., Relativistic Quantum Mechanics(McGraw-Hill, New York, 1964), Sec.~6.4.
\bibitem{messiah} Messiah, A., translated by Temmer, G. M., Quantum Mechanics Vol. II (North-Holland, Amsterdam, 1966), Chap.~XX, Sec.~10.

\end{thebibliography}
\end{document}